\documentclass[conference]{IEEEtran}
\IEEEoverridecommandlockouts

\usepackage{cite}
\usepackage{amsmath,amssymb,amsfonts}
\usepackage{algorithmic}
\usepackage{graphicx}
\usepackage{enumitem}
\usepackage{booktabs}
\usepackage{textcomp}
\usepackage{xcolor}
\usepackage{comment}
\usepackage{multirow}

\def\BibTeX{{\rm B\kern-.05em{\sc i\kern-.025em b}\kern-.08em
    T\kern-.1667em\lower.7ex\hbox{E}\kern-.125emX}}
\begin{document}

\title{Hierarchical Repository-Level Code Summarization for Business Applications Using Local LLMs}

\author{\IEEEauthorblockN{Nilesh Dhulshette, Sapan Shah, Vinay Kulkarni}
\IEEEauthorblockA{TCS Research, Tata Consultancy Services Ltd.,\\
	Pune, India\\
	Email: dhulshette.nilesh1@tcs.com, sapan.hs@tcs.com, vinay.vkulkarni@tcs.com}
}

\maketitle

\begin{abstract}
In large-scale software development, understanding the functionality and intent behind complex codebases is critical for effective development and maintenance. While code summarization has been widely studied, existing methods primarily focus on smaller code units, such as functions, and struggle with larger code artifacts like files and packages. Additionally, current summarization models tend to emphasize low-level implementation details, often overlooking the domain and business context that are crucial for real-world applications. This paper proposes a two-step hierarchical approach for repository-level code summarization, tailored to business applications. First, smaller code units such as functions and variables are identified using syntax analysis and summarized with local LLMs. These summaries are then aggregated to generate higher-level file and package summaries. To ensure the summaries are grounded in business context, we design custom prompts that capture the intended purpose of code artifacts based on the domain and problem context of the business application. We evaluate our approach on a business support system (BSS) for the telecommunications domain, showing that syntax analysis-based hierarchical summarization improves coverage, while business-context grounding enhances the relevance of the generated summaries.
\end{abstract}


\section{Introduction}
In large scale software development projects, it is extremely crucial to have accurate source code comprehension capabilities for effectively developing and maintaining complex software systems. As software projects grow in size and complexity, understanding the functionality, structure, and intent behind the code becomes increasingly difficult. At the same time, the expectation of having well-documented and well-commented code is often unrealistic, especially with tight project deadlines. Even when comments are present, they may not be consistently updated with code revisions. It is reported that developers often spend more than 50\% of their time in comprehending existing code \cite{xia:2018}. Therefore, having an automatic source code summarization capabilities is a critical requirement in large-scale software projects. Code summarization refers to the process of generating concise, human-readable descriptions of various code components such as packages, files, classes, and functions, capturing their purpose and behavior within the business context. High-quality, automatically generated code summaries facilitate tasks such as code understanding, review, maintenance, debugging, and developer onboarding, ultimately leading to improved software development cycles.

Code summarization has been extensively studied, with early research primarily focusing on rule-based or template-based  methods\cite{sridhara:2010,moreno:2013}. In recent years, deep learning approaches have gained significant traction \cite{iyer:2016,hu:2018}, framing code summarization as a machine translation or text summarization problem. These methods typically involve training sequence-to-sequence or transformer-based models. Lately, the accuracy of code summarization has improved significantly with the advent of large language model (LLM) based fine-tuning approaches. However, these models often rely on datasets such as CodeSearchNet \cite{husain:2019} and CodeXGlue \cite{shuai:2021}, which are primarily sourced from GitHub and contain a large proportion of system-level code (at method level). Fine-tuned models trained from such dataset often focus heavily on low-level implementation details and struggle to adequately capture the business context typically found in enterprise software applications. As a result, they do not transfer well to business applications where understanding the underlying intent is just as critical as the implementation details at method level.

Closed-source, API-based LLMs, such as OpenAI's GPT \cite{gpt4}, have significantly advanced code summarization, achieving remarkable improvements not only at the method level but also at the file and package levels. However, privacy concerns limit their adoption, as organizations are still reluctant to share sensitive proprietary source code. While cloud service providers like Azure offer access to these LLMs with data privacy guarantees, organizations still hesitate to adopt them. Local LLMs provide a solution here by allowing deployment on-premises, keeping data private. Despite this advantage, local LLMs still struggle with repository-level summarization, particularly for large files and packages. For example, in our experiments with Llama3.2 (128K context window) \cite{llama32} on a Java file containing 124 functions, many functions were omitted from the file-level summary. While local LLMs ensure privacy and security, their accuracy remains limited for large-scale summarization tasks.

At our organization, we are developing a generative AI-based framework for accelerated delivery of software products in Brownfield setting, with a goal to reduce time and efforts to create bespoke product offerings, managing feature request, and so on. Automatic source code summarization is a critical module in our framework, providing backbone to other coding and software engineering related tasks. With extensive experiments, we discovered that it is difficult to create accurate repository-level source code summarization for business oriented application. 

To address the challenges outlined above, we propose a two-step hierarchical approach for repository-level source code summarization tailored to business applications. While local LLMs may struggle with summarizing entire files or packages, they are effective at summarizing smaller units, such as methods or functions. We exploit this by breaking large code artifacts into smaller components that local LLMs can handle with greater accuracy. Once summaries for these smaller units are generated, we aggregate them to produce file and package level summaries. Specifically, we use a Java parser to generate the abstract syntax tree (AST) of a given Java file and segment it into distinct units, such as class variables, functions, constructors, and enums. Summaries for these individual units are then combined to generate file-level summaries, which are further aggregated into package-level summaries. To ground the summaries in business context, we design custom prompts that elicit the intended purpose of code artifacts based on the domain and problem context of the business application.

We have evaluated our approach on a large proprietary source code for a business support system (BSS) product in the telecommunications domain. However, due to intellectual property constraints, we will present our results from a relatively smaller, publicly available GitHub repository \cite{jtelecom} with similar characteristics\footnote{This repository is also for a BSS product in the telecommunication domain. The repository size is relatively small but well captures the overall idea of a BSS application.}. Our findings demonstrate that both the hierarchical structure and the incorporation of business context significantly improve the accuracy of repository-level source code summarization. The key contributions of this work are:
\begin{enumerate}
  \item A two-step hierarchical approach for repository-level source code summarization using local LLMs (sec.~\ref{sec:approach}), leading to improved coverage (sec.~\ref{sec:filesum_res}).
  \item A novel prompting that grounds code summaries in business applications context (sec.~\ref{sec:domaininfo}), contrasting with vanilla prompting that often focuses on implementation details.
  \item Extensive evaluation on a business support system application in the telecommunication domain, showcasing the effectiveness of our approach (sec.~\ref{sec:setup},~\ref{sec:results}).
\end{enumerate}

\section{Related work}
Code summarization has a long history of research, with early works primarily focusing on pattern-based and template-based models \cite{sridhara:2010,moreno:2013}. With the advent of deep learning, the field shifted toward neural machine translation models, such as sequence-to-sequence summarization for languages like C\# \cite{iyer:2016} and Java \cite{hu:2018}. This progression was soon followed by transformer-based models, including encoder-decoder architectures \cite{ahmad:2020} and BERT-based approaches \cite{wang:2020}. More recently, the emergence of code-specific LLMs, such as CodeLlama \cite{codellama:2024}, StarCoder \cite{starcoder2:2024}, and DeepSeek-Coder \cite{deepseek:2024}, has significantly improved summarization accuracy. These advances are also attributed to the availability of large-scale datasets like CodeSearchNet \cite{husain:2019} and the code-text subset of CodeXGLUE \cite{shuai:2021}. However, existing approaches have primarily focused on smaller code units, such as individual functions or methods, and often struggle to scale to larger code structures, such as entire files or packages. As a result, their performance diminishes when applied to larger codebases that require a broader, more comprehensive understanding. 

Recently, researchers have started exploring the utility of LLMs in repository-level software engineering tasks such as requirement generation \cite{ren:2024,xie:2023}, code generation \cite{nguyen:2024,chen:2023}, program repair \cite{hossain:2024,li:2024}, and commit message generation \cite{jung:2021,xue:2024}. For code summarization, Rukmono et. al. \cite{rukmono:2023} have focused on component-level summaries by leveraging abstract syntax trees and using a chain-of-thought prompting technique with LLMs. In contrast, P-CodeSum \cite{yun:2024} creates artifact-specific prompts by using project-specific few-shot \textit{\{code, summary\}} examples with a neural prompt selector, allowing it to better capture the unique nuances of a given codebase. Despite these advancements, the challenge of summarizing high-level code artifacts such as entire files or packages, remains relatively under explored in the literature.  This gap is particularly important for large-scale projects, where summarizing individual functions may no longer provide a sufficient understanding of the codebase as a whole.

A key challenge in repository-level summarization lies in the nature of existing models and datasets, which tend to emphasize low-level implementation details. These models typically focus on \textit{what} the code does and \textit{how} it achieves that \cite{fangwen:2023,chen:2021}, often emphasizing identifier names and control flow. While this approach works well for simpler, isolated code units, it overlooks the domain-specific aspects and business context. Real-world software systems, however, are typically built to solve particular business problems within specific domains, such as customer relationship management in the telecommunications domain or digital twins in the parcel delivery domain. Current models often neglect this broader context, which hinders their ability to transfer effectively to business applications where understanding the intent behind the code is crucial for meaningful summarization. To address this gap, we propose a novel code summarization approach that not only generates repository-level summaries but also grounds them in business context.

\section{Repository-level Code Summarization}\label{sec:approach}

Our goal is to develop a repository-level code summarization system that captures both low-level code artifacts (such as variables and functions) and the intent of higher-level artifacts (such as files and packages). While local LLMs are effective for summarizing smaller, function-level code, our observations show that they still struggle with accurate and coherent repository-level summaries.

\begin{figure*}[h]
	\centering
	\includegraphics[width=1\textwidth]{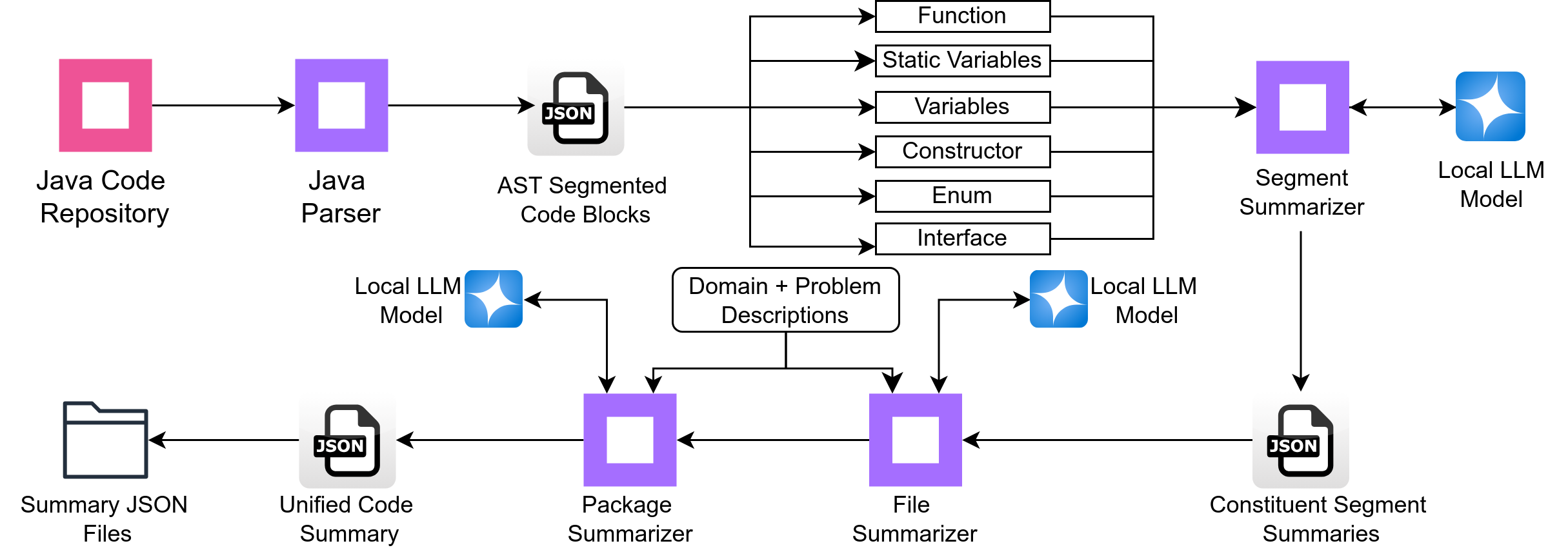} 
	\caption{Two-step hierarchical approach to repository-level code summarization tailored for business applications}
	\label{fig:Workflow}
\end{figure*}

To address this issue, it is essential to break down repository-level artifacts into smaller, manageable units that local LLMs can process effectively. By generating cohesive summaries at the block level, we can then combine them to form a comprehensive understanding at the repository level. Figure~\ref{fig:Workflow} illustrates this approach. First, we decompose individual files in the repository using a language parser to extract meaningful file-level segments. Next, we apply custom summarizers, tailored to each segment type, to generate detailed summaries. These segment-level summaries are aggregated into file-level summaries, incorporating domain and business context, which are further combined into package-level summaries. All summarization steps in this process are implemented using LLMs in inference-only mode. The rest of this section describes the key components of our system.

\subsection{Code Segmentation using Abstract Syntax Tree}\label{ast}
To effectively segment a codebase, it is crucial to identify cohesive components that can be interpreted independently. Ideally, one would like each individual source file to be treated as a distinct component, allowing its summary to be easily combined into package and repository level summaries. However, this approach becomes problematic when the size of a single file grows significantly. In such cases, summarizing large source files may result in the omission of important details, such as individual function descriptions, or result in an oversimplified summary that neglects key portions of the original code. Therefore, it is essential to decompose large source files into smaller, coherent, and meaningful units to enable more detailed summarization.

To achieve this, we perform syntactic analysis on the source file using a Java parser \cite{javalang} to generate its Abstract Syntax Tree (AST), which provides an intermediate representation of the code. The AST represents the corresponding source file as a hierarchical tree, with nodes corresponding to various language constructs, such as declarations, expressions, definitions, statements, etc. We then traverse the AST to identify nodes whose type matches one of the following categories: Function, Variable, Constructor, Enum, and Interface. For each such node, we consider the subtree rooted at that node as a logical unit and serialize it into a code segment. Additionally, we retain localization information, including the file name and the start and end line numbers of the segment within the file.

\subsection{Segment-level Code Summarization}
The individual segment types discussed in the previous section each play distinct roles in contributing to the overall understanding of a source file. For instance, variables generally serve as value holders, while static variables provide constant values with specific purposes. Similarly, functions typically implement specific features, while certain types of functions, such as constructors, are primarily used for initializing variables. The key to successful summarization lies in understanding the unique roles these segments play in constructing a comprehensive understanding of the source file. To achieve this, we design custom prompts for each segment type, focusing on their role and purpose within the code. The following outlines the key aspects considered for each segment type:
\begin{enumerate}
\item \textbf{Function:} For functions, we consider six key aspects: name, input, output, workflow, side effects, and purpose. The function name is extracted along with its input arguments, whose meanings are inferred from surrounding context, such as the function name and argument type. Similarly, the output is inferred based on its type and usage. The workflow captures a comprehensive, line-by-line summary of all statements within the function. Side effects focus on global updates, state changes, and logging information. All of these aspects are then used to derive the function's primary purpose.
\item \textbf{Constructor:} For constructors, our goal is to identify the key properties that define class creation, such as instance variables and initialization values, helping us understand the initial state of objects and their intended behavior.
\item \textbf{Variable:} For variables, we aim to capture their type, scope, and role in the code. Static variables are treated similarly, with a focus on their constant values, which may have a broader impact across the application.
\item \textbf{Enum:} For enums, we aim to identify the constant values they represent and the role of these constants in controlling application logic.
\item \textbf{Interface:} For interfaces, we examine their role in defining a contract for classes to implement with a focus on the intended purpose of methods. 
\end{enumerate}

The exact prompts used by our system are detailed in the experiments section.

\subsection{Repository-level Code Summarization}
Once the summaries for individual segments are generated, we first combine them to generate file-level summary. Similarly, the file-level summaries are then combined to generate package-level summaries. 
\begin{enumerate}
    \item \textbf{File-level summary:} To obtain a file-level summary, we first order its segments according to their positions using the localization information. We then combine them using a custom prompt that not only uses comprehensive segment-level summaries but also considers the domain and the problem context (as detailed in sec.~\ref{sec:domaininfo}) of the business application to determine the file's \textit{purpose} and \textit{role} within the overall codebase, and its \textit{key functionality}.
    \item \textbf{Package-level summary:} We combine the individual file-level summaries to obtain a package-level summary. Since understanding the \textit{purpose} and \textit{role} of each individual file provides the necessary context for generating the package summary, we use only the file summary and omit all the details of its constituent segment summaries. The package-level summaries are then similarly combined to obtain the repository-level summary.
\end{enumerate}

\subsection{Grounding in Business Application}\label{sec:domaininfo}
LLMs, trained on vast amounts of data from the web, capture a broad range of knowledge about real-world and abstract entities, their properties, and their relationships \cite{petroni:2019}. They have also been shown to contain domain-specific knowledge. For instance, knowledge about gene, protein, etc. in Biomedical domain. While LLMs can handle generic tasks out-of-the-box, their performance can be significantly improved when they are grounded in the specific domain and problem context of a business application. We have observed this across various domains, including enterprise digital twins, manufacturing, and chemical synthesis.

In the context of code summarization, grounding the LLM to the business domain and problem context is essential for generating accurate, meaningful summaries.
\begin{enumerate} 
    \item \textbf{Grounding to Domain:} Providing a succinct description of the domain helps the LLM develop a deeper understanding of the concepts relevant to the business application. This understanding allows the LLM to generate summaries that align more closely with the domain-specific context. 
    \item \textbf{Grounding to Problem Context:} Similarly, offering detailed context about the specific problem being addressed enhances the LLM’s understanding of the task at hand. This enables the LLM to uncover the intended purpose of various source code artifacts, improving the quality of the summaries.
\end{enumerate}

For the experiments in this paper, we focus on grounding the LLM in the \textit{\textbf{telecommunication}} domain, with the problem context being a \textit{\textbf{business support system}}. Details of this grounding are described further in the experiments section.

\section{Experimental Setup}\label{sec:setup}
Our organization has developed an operations and business support system (OSS/BSS) product for the telecommunications domain. While we have evaluated our approach on the full product codebase, this paper presents our findings based on a smaller, publicly available GitHub repository \cite{jtelecom} that mirrors the characteristics of our BSS module. This repository includes code for supporting various business functions such as sales and service management, customer management, and billing. The codebase is written using the Java Spring framework and consists of 122 Java class files, 20 interfaces, 762 functions, 704 variables (including member fields and statics), and 11 enums.

We experimented with instruct-tuned general and code-specific LLMs, all in inference-only mode. The following LLMs were used in our experiments:
\begin{enumerate}
    \item \textbf{Llama-3}: An instruct-tuned variant of Llama-3 (8B) that performs well on both general and code-related tasks \cite{llama32}.
    \item \textbf{Starchat2}: An instruct-tuned variant of Starcoder-2 (15B) \cite{starchat2}, optimized for code-related tasks.
    \item \textbf{Codestral}: A code-specific 22B LLM from Mistral \cite{codestral}, designed for code generation tasks.
\end{enumerate}

\subsection{Ground-Truth data using GPT-4}
To obtain ground-truth data for evaluation, we utilized \mbox{GPT-4} \cite{gpt4} from OpenAI, known for its strong performance in tasks ranging from language generation to code understanding. We randomly selected 10 files from our repository, covering multiple packages, to design prompts for generating file-level summaries. The following outlines the process we followed:
\begin{enumerate}
    \item \textbf{Segment Extraction:} We first extracted all relevant code segments such as functions, variables, etc., from the selected files using syntax analysis.
    \item \textbf{Prompt Optimization:} We then iteratively optimized segment-specific prompts, using techniques like in-context learning and chain-of-thought prompting, incorporating feedback from subject matter experts (SMEs).
    \item \textbf{File-Level Summaries:} Using the segment-level summaries, we generated file-level summaries through an additional round of prompt engineering.
\end{enumerate}
Next, we used the optimized prompts to generate summaries for all files in the repository. We then selected two packages and iteratively developed prompts for package-level summarization based on the files within the packages. Finally, the optimized package-level prompt was applied to generate summaries for all the remaining 34 packages, resulting in ground-truth data for repository-level summarization. It should be noted that, rather than relying solely on GPT-4 output, the above process was guided by detailed feedback and validation from SMEs, resulting in reliable ground-truth data.

In summary, using \mbox{GPT-4}, we obtained ground-truth summaries for the following code elements: functions, constructors, variables, static variables, enums, interfaces, classes (files), and packages.

\subsection{Evaluation Metrics}
We evaluate our results using traditional summarization metrics such as ROUGE-L \cite{rouge} and BLEU \cite{bleu}, transformer-based metrics like BERTScore \cite{bertscore} and semantic similarity, as well as the LLM-As-A-Judge model in G-Eval \cite{geval}. With G-Eval, we consider the following criteria:- \textit{Completeness:} the summary should cover all aspects of the code; \textit{Conciseness:} the summary should be concise; \textit{Correctness:} the summary should not hallucinate; \textit{Cohesiveness:} the summary should be cohesive; and \textit{Domain Specificity:} the summary should  reflect domain-specific terms and concepts.

\section{Results and Analysis}\label{sec:results}
\subsection{Segment-level Summarization}
This section begins with prompts and results for function summarization, followed by results for other segment types.

We designed the prompts for function summarization progressively, starting with simple instructions and gradually incorporating more detailed prompts that leverage chain-of-thought reasoning and in-context learning. The following describes these prompts:

\begin{enumerate}
    \item \textbf{Generic Prompt:} This prompt provides a general instruction to summarize the code segment, with the expectation to obtain important aspects such as purpose, arguments and return types.
    \item \textbf{Structured Prompt:} This prompt explicitly asks the LLM to generate structured output with clearly defined fields including purpose, inputs, outputs, workflow, side effects, and a final summary. By explicitly specifying these fields, the prompt guides the LLM in focusing on the most relevant aspects of the code, improving the accuracy and clarity of the output.
    \item \textbf{Structured Prompt+1S:} This version is an in-context learning variation of the structured prompt, where the LLM is provided with a single example function and its summary. This one-shot example helps the model understand the expected format and the level of detail required for each field.
\end{enumerate}
Table~\ref{tab:function_prompts} provides the exact prompts\footnote{These prompts are designed keeping Brownfield software development use cases in focus. They might need to be modified for other use cases.} used in our system for function summarization.

\begin{table}[!tbp]
	\centering
	\caption{Prompts for Function Summarization}
	\begin{tabular}{|p{0.95\columnwidth}|}
		\hline
		\textbf{Generic Prompt:} \\ 
		\hline
		You are an expert Java code assistant with a deep understanding of Java programming. Your task is to analyze the provided Java function and create a comprehensive summary that captures all its essential aspects.\\
		\hline
		\hline
		\textbf{Structured Prompt: Generic Prompt followed by}\\
		\hline
		The summary should include the following components:\\
		Function Name: The name of the function.\\
		Inputs: The parameters and their types.\\
		Outputs: The return type and value.\\
		Purpose: A brief description of what the function does and its purpose.\\
		Workflow: An outline of the logic and steps involved in the function.\\
		Side Effects: Any potential side effects that the function may have.\\
		Final Summary: A concise overview of the function’s overall functionality.\\
		\hline
		\hline
		\textbf{Structured Prompt+1S:}\\
		\hline
		Structured Prompt + \{function and its summary with field level details\}\\
		\hline
	\end{tabular}
	\label{tab:function_prompts}
\end{table}
\begin{table}[!tbp]
	\centering
	\caption{Summarization prompts for constructor, variable (member field and static), interface, and enum}
	\begin{tabular}{|p{0.95\columnwidth}|}
		\hline
		\textbf{Constructor} \\ 
		\hline
		You are an expert Java code assistant who is great at explaining any java code snippet. You will be provided with constructor of a java class. Your task is to summarize it in the below format.
		\begin{itemize}[noitemsep,topsep=0pt]
			\item Name:
			\item Parameters:
			\item Member variables initialized:
		\end{itemize}
		\{One-shot example\}\\
		\hline \hline
		\textbf{Variable}\\
		\hline
		You are an expert Java code assistant who is great at explaining any java code snippet. You will be provided with a Java class member variable which may contain the value it is initialized with. Your task is to explain the value. If it relates to an SQL query, explain the purpose of the query. Please refrain from disclosing the actual value or the method of its implementation in your explanation.\\
		\{Few-shot examples\}\\
		\hline \hline
		\textbf{Interface}\\
		\hline
		You are an expert Java code assistant. Your task is to analyze a provided Java interface. For the given interface, please do the following:
		\begin{itemize}[noitemsep,topsep=0pt]
			\item List all function declarations, including their signatures and explain them
			\item Identify any variable declarations
		\end{itemize}\vspace*{-\baselineskip}\\
		\hline \hline
		\textbf{Enum}\\
		\hline
		You are an expert Java code assistant. Your task is to analyze a provided Java Enum. For the given Enum, please do the following:
		\begin{itemize}[noitemsep,topsep=0pt,]
			\item List all the constants declared in the Enum
			\item Describe any additional functionality that this Enum provides
			\item Explain the probable purpose of the Enum
		\end{itemize}\vspace*{-\baselineskip} \\ 
		\hline
	\end{tabular}
	\label{tab:segment_prompts}
\end{table}

\begin{table*}[!htbp]
   \centering
   \caption{Function Summarization: chain-of-thought with in-context learning in Structured+1S gives the best summarization results (SS: Semantic Similarity between sentence embeddings)}
      \begin{tabular}{p{0.07\textwidth}|p{0.11\textwidth}|cccc|cccc}
      \hline
      \textbf{Model} & \textbf{Prompting} & \textbf{ROUGE-L} & \textbf{BLEU} & \textbf{BERTScore} & \textbf{SS} & \textbf{Correctness}  & \textbf{Conciseness} & \textbf{Completeness} & \textbf{Cohesiveness} \\
      \hline
      \multirow{3}{0.07\textwidth}{Llama-3} & Generic & 0.26 & 4.36 & 0.65 & 0.91 & 0.752 & 0.700 & 0.739 & 0.733 \\
      & Structured & 0.45 & 11.86 & 0.72 & 0.94 & 0.796 & 0.744 & 0.891 & 0.831 \\
      & Structured+1S & 0.587 & 17.38 & 0.76 & 0.95 & 0.814 & 0.738 & 0.887 & 0.839 \\
      \hline
      \multirow{3} {0.07\textwidth}{Starchat2} & Generic & 0.24 & 3.04 & 0.63 & 0.91 & 0.745 & 0.704 & 0.749 & 0.749 \\
      & Structured & 0.44 & 10.32 & 0.74 & 0.96 & 0.778 & 0.721 & 0.870 & 0.816 \\
      & Structured+1S & 0.542 & 20.40 & 0.80 & 0.97 & 0.785 & 0.689 & 0.876 & 0.817 \\
      \hline
      \multirow{3}{0.07\textwidth}{Codestral} & Generic & 0.30 & 8.39 & 0.67 & 0.934 & 0.778 & 0.636 & 0.803 & 0.775 \\
      & Structured & 0.42 & 11.46 & 0.721 & 0.95 & 0.824 & 0.654 & 0.906 & 0.817 \\
      & Structured+1S & 0.577 & 21.26 & 0.81 & 0.97 & 0.840 & 0.645 & 0.920 & 0.807 \\
      \hline
      \end{tabular}
      \label{tab:functionresults}
\end{table*}
Table~\ref{tab:functionresults} reports the results for function summarization. The baseline generic prompt performs reasonably well across different LLMs. As expected, directing the LLMs to generate structured outputs using the chain-of-thought prompt enhances accuracy compared to the baseline. Additionally, incorporating in-context learning with a one-shot example leads to significant improvements. On average, the structured one-shot prompt outperforms the generic prompt by more than 13\% in completeness, 5\% in correctness and cohesiveness, and approximately 1\% in conciseness. These results suggest that while LLMs tend to produce concise summaries, they may struggle with completeness, correctness, and cohesiveness. Both chain-of-thought prompting and in-context learning contribute substantially in improving summaries in these areas.

Similar to function summarization, we applied generic and structured prompts, incorporating chain-of-thought and in-context learning, to optimize summarization for other segment types, including constructors, variables, enums, and interfaces. For each segment, we tailored the prompts to highlight the specific aspects they represent and their role within the broader codebase. For example, static variables often store key state information or configuration values that influence the behavior of a class. In contrast, enums and their constants define fixed sets of values, helping to constrain the potential values of certain variables. Understanding constructor skeletons is crucial, as they reveal mandatory member variables and provide insights into class functionality. Table~\ref{tab:segment_prompts} describes the final prompts used for each segment type.

\begin{table}[!tbp]
    \centering
    \caption{Summarization results for other segments (SS: Semantic Similarity)}
    \resizebox{0.5\textwidth}{!}{ 
        \begin{tabular}{p{0.11\columnwidth}lcccc}
            \hline
            Segment & Model & ROUGE-L & BLEU & SS & BERTScore \\ 
            \hline
            \multirow{2}{0.11\columnwidth}{Constructor} & LLama-3  & 0.77 & 20.58 & 0.96 & 0.88  \\ 
            & Starchat2 & 0.64 & 15.69 & 0.93 & 0.83 \\ 
            \hline
            \multirow{2}{0.11\columnwidth}{Variable} & LLama-3 & 0.40 & 14.30 & 0.93 & 0.73  \\ 
            & Starchat2 & 0.41 & 14.99 & 0.94 & 0.73 \\ 
            \hline
            \multirow{2}{0.11\columnwidth}{Interface} & LLama-3  & 0.49 & 15.39 & 0.96 & 0.78  \\
            & Starchat2 & 0.44 & 11.44 & 0.93 & 0.73 \\
            \hline
            \multirow{2}{0.11\columnwidth}{Enum} & LLama-3 & 0.43 & 12.73 & 0.97 & 0.75  \\ 
            & Starchat2 & 0.37 & 6.32 & 0.94 & 0.67 \\
            \hline            
        \end{tabular}
    }
    \label{tab:results_other_segs}
\end{table}
The summarization results, as evaluated using the traditional metrics, are reported in Table~\ref{tab:results_other_segs}. The results seem promising across all segment types, with Llama-3 achieving the highest summarization quality. Constructors, which typically exhibit a limited number of variations, achieve particularly high scores. In contrast, variables and enums, due to their diverse initializations, exhibit greater variability, resulting in slightly lower scores. Interfaces, being collections of abstract methods, are not segmented further but instead processed directly at the file level. This approach likely explains their relatively lower scores. However, our manual inspection confirms that the summarization quality remains generally consistent and satisfactory across all segment types.

\subsection{File Level Summarization}\label{sec:filesum_res}
To generate file-level summaries, we first combine the segment-level summaries while considering the sequential order of the segments. We then generate the \textit{purpose} of the given file, eliciting the \textit{role} it plays in the overall repository and its \textit{key functionality}. As mentioned earlier, LLMs, trained on vast amounts of data from the web, already carry a high-level understanding of various domains and perform well out-of-the-box on many NLP tasks, including code summarization. However, providing the right context, both in terms of the domain at hand and the problem context of the business application, helps LLMs in improving the generated summaries, especially for components that require a deeper understanding, such as the intent or purpose of the entire file. Table~\ref{tab:domainprompt} outlines these additional descriptions for the BSS application in the telecommunications domain, which play a key role in the file-level summarization prompt shown in table~\ref{tab:filepackage_prompt}.

\begin{table}[!tbp]
    \centering
    \caption{Descriptions for grounding LLMs to business application context}
    \begin{tabular}{|p{0.95\columnwidth}|}
        \hline
        \textbf{Domain Description:} \\ 
        \hline
        You specialize in the telecommunication domain, which includes technologies, services and, infrastructure that transmit information over distances in various forms, such as voice, data, and text. It includes various services, such as mobile networks, broadband internet, and$\;$...$\;$ Key components include network operators, equipment manufacturers, and service providers,$\;$...$\;$ The domain also involves a range of business functions, including customer relationship management, billing, service provisioning, and network optimization\\
        \hline \hline
        \textbf{Problem Context Description:} \\ 
        \hline
        The code repository contains software for the Business Support System (BSS) application. It includes various user roles, such as Residential Client, Business Client, Administrator, CSR (Customer Service Representative), and Problem Management Group, each with specific permissions and capabilities. Users can log in via email to access$\;$...$\;$ Additionally, the software supports report generation, graph creation, data searching and sorting, and email responses.\\
        \hline
    \end{tabular}
    \label{tab:domainprompt}
\end{table}
\begin{table}[!tbp]
	\centering
	\caption{File-level and Package-level Summarization Prompts}
	\begin{tabular}{|p{0.95\columnwidth}|}
		\hline
		\textbf{File Summarization Prompt} \\ 
		\hline
		You are an expert Java code assistant with a deep understanding of Java programming.\\
		\texttt{Domain Description + Problem Context Description} \\
		The source Java file is converted into its textual description and includes package name, import statements, class/enum/interface name, member field description, constructor summary, and a list of functions with summaries.\\
		Generate a file-level summary based on the textual description of a source Java file. This summary should articulate what the class aims to achieve and the role it plays in the repository. While generating summaries, consider the aforementioned domain and problem description. The summary should be concise.\\
		\textbf{Output Format:}\\ 
		The output should include \lq Role \rq, \lq Key functionality\rq, and \lq Purpose \rq.\\
		\hline \hline
		\textbf{Package Summarization Prompt:} \\ 
		\hline
		You are an expert Java code assistant specializing in telecommunication industry. You have been provided with summaries of individual Java files within a package. Based on these summaries, your task is to:
		\begin{itemize}
			\item Generate the overall purpose of the package, explaining its role within the repository.
			\item Describe how the package achieves its goals by detailing the key functionalities and interactions between the files in the package.
		\end{itemize}\\
		\hline
	\end{tabular}
	\label{tab:filepackage_prompt}
\end{table}

The LLMs we used for segment-level summarization encountered limitations when applied to file-level summarization, mainly due to their limited context window. To address this, we focused on the recently proposed Llama-3.2 model, which supports a context length of 128K tokens, to generate the file-level summary. We experimented with two variants: Llama-3.2 where segment summaries were obtained using the Llama-3 model; and Llama-3.2$^\#$ where segment summaries from the Starchat2 model were used. As shown in Table~\ref{tab:res_filepurpose}, grounding the model with both domain and problem context significantly improved domain relevance (DS) by over 7\%. Importantly, the conciseness and cohesiveness of the generated summaries remained consistent with the baseline, where no additional business context was provided. Moreover, grounding also led to improvements in the completeness of the summaries. These findings confirm that grounding LLMs in the specific business application context enhances their ability to generate better, more relevant file-level summaries. Table~\ref{tab:example} illustrates the generated summary for a sample Java file from our repository.

\begin{table}[!tbp]
    \centering
    \caption{File-level Summarization: Domain and Problem context grounding generates better summaries (DS:domain specificity)}
        \begin{tabular}{p{0.16\columnwidth}p{0.11\columnwidth}cccc}
            \hline
            \textbf{Models} & \textbf{Grounding} & \textbf{DS} & \textbf{Concise} & \textbf{Complete}  & \textbf{Cohesive}  \\ 
            \hline \hline
            \multirow{2}{0.16\columnwidth}{Llama-3.2} &  With & 0.637 & 0.662 & 0.904 & 0.844  \\
            &  Without & 0.568 & 0.672 & 0.868 & 0.832 \\            
            \hline
            \multirow{2}{0.16\columnwidth}{Llama-3.2$^\#$} &  With & 0.66 & 0.684 & 0.872 & 0.815\\
            & Without & 0.614 & 0.669 & 0.865 & 0.825  \\            
            \hline 
        \end{tabular}
    \label{tab:res_filepurpose}
\end{table}

\subsection*{Evaluating Summarization Coverage}

Prompting LLMs directly with large source files often leads to missing information in the generated summaries. For example, all the LLMs we tested struggled to reliably summarize many functions in a Java file containing 124 functions. To test this quantitatively, we provided complete source files as input to the LLMs and evaluated the generated summaries for coverage, focusing on different segment types. For instance, we considered a function to be faithfully represented in the summary only if it included the purpose of that function (even partially). As we couldn't identify consistent patterns to automatically verify this, we manually inspected all the files in the repository. Our tests with Llama-3.2 showed that direct file-level summarization missed approximately 11\% of the functions and 24\% of the variables. A similar trend was observed with GPT-4, where 9\% of functions and 11\% of variables were omitted from the summaries. In contrast, our approach avoids this issue, as it is explicitly designed to generate summaries for all segments, ensuring that every relevant component is included in the file-level summaries.

\begin{table}[!tbp]
	\centering
	\caption{Example Summary for a FillProductPricesController.java}
	\begin{tabular}{|p{0.95\columnwidth}|}
		\hline
		The \lq FillProductPricesController\rq class is a Spring-based controller that handles requests related to filling product prices by region. Its primary purpose is to manage the pricing process for residential customers, providing a user-friendly interface for administrators to fill in prices. \\
		\textbf{Role}: This controller plays a crucial role in the cart services master service application by providing a centralized interface for administrators to manage product prices by region. It ensures that prices are filled correctly for residential customers, enabling the application to provide accurate pricing information to customers.\\
		\textbf{Key Functionality:} The controller offers two primary functions:
		\begin{enumerate}[noitemsep,topsep=0pt,wide, labelwidth=!, labelindent=0pt]
			\item Filling product prices by region: This function retrieves a list of places for filling in tariff by region and sets the model and view data accordingly.
			\item Retrieving a page of products filled by price by regions: This function handles a GET request to the \lq fillProductPrices\rq endpoint, checks that the product customer type is \lq Residential \rq, and returns a \lq ModelAndView \rq object containing all \lq Place \rq model attributes.
		\end{enumerate}
		\textbf{Purpose:} The \lq FillProductPricesController \rq class is designed to simplify the pricing process for residential customers, ensuring that prices are filled correctly and providing a seamless user experience for administrators. Its primary intent is to facilitate accurate pricing information and efficient management of product prices by region.\\
		\hline
	\end{tabular}
	\label{tab:example}
\end{table}

\subsection{Package-level Summarization}
We generate a comprehensive understanding of the overall functionality and the role of each package in the repository by using domain and problem context-aware file-level summaries as input. Table~\ref{tab:filepackage_prompt} presents the prompt for package-level summarization. Our repository contains 36 packages in total. The generated summaries were manually evaluated by expert software developers\footnote{The summaries were evaluated individually by three experts, and their feedback was consolidated for overall evaluation.}
experienced in the telecommunication industry. The experts found the package summaries to be highly informative and well-aligned with the problem domain, covering essential details that help understand each package's functionality within the larger context.

\section{Conclusions and Future Work}
In this paper, we presented a two-step hierarchical approach for repository-level source code summarization, specifically tailored to business applications. Our approach first decomposes large code artifacts into smaller segments, enabling local LLMs to summarize them with high accuracy. These segment-level summaries are then aggregated to produce coherent file and package-level summaries. To enhance the relevance and quality of the summaries, we incorporate domain and problem descriptions to ground them in the business context. Our evaluation on a publicly available repository for a business support system in the telecommunications domain demonstrates the effectiveness of our approach, yielding concise, coherent, and domain-aware summaries that capture all key constructs in the repository.

For future work, we plan to incorporate agentic models with self-reflection capabilities to further enhance the summarization quality. We also intend to extend our approach to a wider range of domains, such as healthcare and finance, to assess its adaptability across diverse business contexts. Another promising direction is exploring multi-modal LLMs, where we integrate image artifacts, such as class or activity diagrams, with code to enhance repository-level code comprehension.

\bibliographystyle{IEEEtran}
\bibliography{llm4code_code_summarization_2025}

\end{document}